\documentclass[epj,reprint,showpacs,twocolumn,superscriptaddress]{revtex4}
\usepackage{amsmath}
\usepackage{graphicx}
\usepackage{bm}
\usepackage{epsfig}
\usepackage{natbib}

\begin{document}

\title{Deceleration and trapping of heavy diatomic molecules using a ring-decelerator}

\author{Joost E. van den Berg, Samuel Hoekman Turkesteen, Eric B. Prinsen, and Steven Hoekstra}
\affiliation{Kernfysisch Versneller Instituut, Zernikelaan 25, 9747 AA, Groningen, The Netherlands. Tel: +(31)50 363 9713; E-mail: s.hoekstra@rug.nl}

\begin{abstract}

We present an analysis of the deceleration and trapping of heavy diatomic molecules in low-field seeking states by a moving electric potential. This moving potential is created by a ``ring-decelerator'', which consists of a series of ring-shaped electrodes to which oscillating high voltages are applied. Particle trajectory simulations have been used to analyze the deceleration and trapping efficiency for a group of molecules that is of special interest for precision measurements of fundamental discrete symmetries. For the typical case of the SrF molecule in the $(\mathrm{N,M})=(2,0)$ state, the ring-decelerator is shown to outperform traditional and alternate-gradient Stark decelerators by at least an order of magnitude. If further cooled by a stage of laser cooling, the decelerated molecules allow for a sensitivity gain in a parity violation measurement, compared to a cryogenic molecular beam experiment, of almost two orders of magnitude.
\end{abstract}
\maketitle

\section{Introduction}
\subsection{Motivation}
Precision spectroscopy of atoms and molecules has been a powerful tool to test physics theory for many decades. The remarkable recent advances in the techniques to control simple molecules are pushing the frontier even further~\cite{Doyle:2004vk, Carr:2009ch, Dulieu:2009ei}. Molecular spectroscopy can now be used to test the fundamental discrete symmetries that are at the basis of the Standard Model of particle physics. A prime example are the experiments \cite{Hudson:2002ia,Kozlov:2002cz,ShaferRay:2006fy,Vutha:2010ju,Hudson:2011hs} that search for an electric dipole moment of the electron (eEDM) by exploiting the huge internal electric field of a molecule. The sensitivity that can be reached in these measurements is intimately linked to the experimental control of the molecules. Possible improvements can be made by using more sensitive molecular systems, more intense sources of molecules with better state preparation, and by an increase of the interaction time. By using traps instead of molecular beams, the interaction time can potentially be increased by a factor of a few hundred~\cite{Tarbutt:2009cf}. At our institute we have therefore started a research program towards the deceleration and trapping of molecules that have a large inherent sensitivity for tests of fundamental discrete symmetries. Specifically, we aim to decelerate and trap SrF molecules, by adapting the recently developed ring-decelerator technique~\cite{2010PhRvA..81e1401O, Meek:2011bpa}, as a highly sensitive probe of nuclear-spin-dependent parity violation.

It has been known for a long time that heavy diatomic molecules offer a high sensitivity to study the violation of discrete symmetries such a parity violation\cite{Sushkov:1978vj,Flambaum:1985fw,Kozlov:1995ce,Demille:2008he,2010PhRvA..82e2521I}. None of the suitable diatomic molecules, such as YbF, SrF, PbF and RaF have however been trapped. Stark deceleration\cite{Bethlem:1999wa} and subsequent trapping has been demonstrated for a number of molecules in recent years~\cite{Bethlem:2002uc,Meerakker:2005vh,Hoekstra:2007vx,Hoekstra:2007dk,Sawyer:2007ga,Gilijamse:2007wd,Stuhl:2008hd}. These molecules are relatively light ($< 25$ atomic mass units (amu)) and are therefore not sensitive to parity violation or the electron EDM. A second possible approach, that has resulted in the coldest samples of trapped molecules so far, is the association of ultracold atoms into ultracold molecules\cite{KNi:2008hm,Ospelkaus:2010iq}. This approach is unfortunately not feasible for the fluorine containing alkaline-earth radicals. A third approach is to use buffer gas sources as slow and intense beams of heavy diatomics, which have recently been developed with the aim of precision measurements\cite{Skoff:2009ek,2011PCCP...1318936B}. For an overview of other possible techniques to cool molecules we refer to a number of review articles~\cite{Doyle:2004vk,Carr:2009ch,Dulieu:2009ei}. For the remainder of this article we focus on Stark deceleration.

\subsection{Stark deceleration of heavy molecules}
\begin{figure}
\resizebox{\linewidth}{!}{\includegraphics{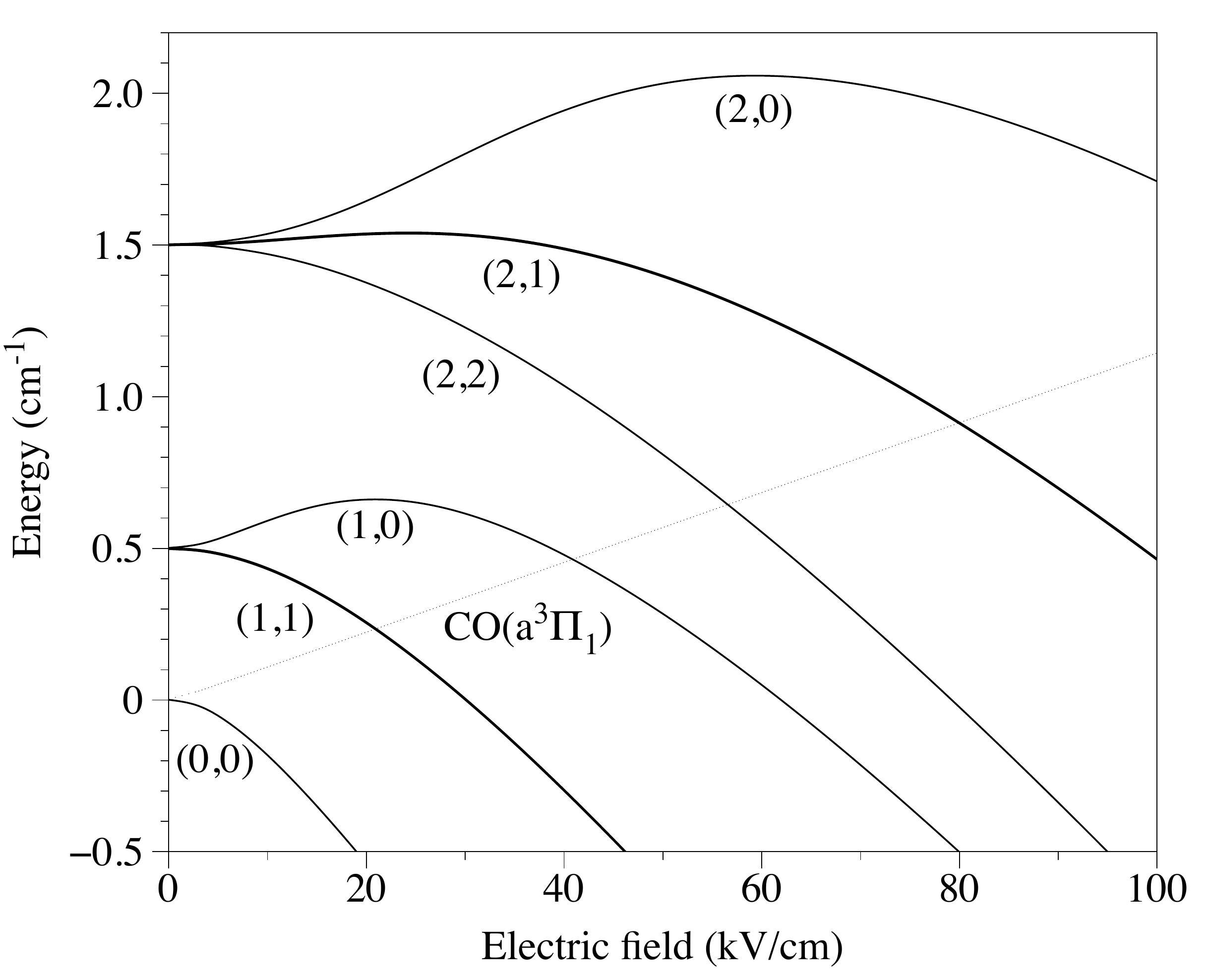}}
  \caption{\label{starkcurves}The Stark shift of the three lowest rotational levels of the $^{88}$SrF molecule in its $\mathrm{X}^{2}\Sigma^{+}$ groundstate. The levels are labelled $(N,M)$ where $N$ is the rotational quantum number and $M$ the projection of $N$ on the electric field axis. Hyperfine structure is not visible on this level. For comparison also the Stark shift of $M_{J}\Omega=-1$ component of the $J=1$ level of CO in the $\mathrm{a}^{3}\Pi_{1}$ state is shown (dotted line).}
\end{figure}

The Stark deceleration of the heavier molecules (which can be roughly defined as having a mass $>$ 100 amu) brings two challenges. First, for a fixed initial velocity and deceleration strength the large mass requires a rather long decelerator, which puts high demands on the stability of the deceleration process. Second, because of the small spacing of rotational levels, at modest electric fields (20-30 kV/cm) the slope of the Stark curve of the lowest rotational levels turns negative; at higher fields these states are all so-called high-field seekers, and are attracted to the high field at the electrodes. At the turning point the maximum Stark shift is for the lowest rotational states only a fraction of a wavenumber. This is illustrated in Fig.~\ref{starkcurves} for the lowest rotational states of SrF in its $\mathrm{X}^{2}\Sigma^{+}(v=0)$ electronic and vibrational ground state. For comparison, also the Stark shift of the CO molecule in the metastable $\mathrm{a}^{3}\Pi_{1}$ state is shown.

To address the challenge of the deceleration of heavier molecules a number of approaches has already been taken. The deceleration of molecules in a high-field seeking state (with maximum energy at minimum electric field) is possible using an Alternate Gradient (AG) decelerator. From a number of proof-of-principle experiments that have been performed~\cite{Bethlem:2002ed,Tarbutt:2004du,Bethlem:2006vx} it has become clear that the stability of such a decelerator depends very critically on the alignment of the electrodes, causing the efficiency to drop rapidly with decelerator length. Another very critical part would be the transfer from an AG decelerator to an AC trap that can trap the high-field seeking molecules~\cite{Bethlem:2006vx}.

Alternatively, the molecules could be decelerated in the weak-field seeking part of an excited rotational state~\cite{Tarbutt:2009cf}, such as the SrF(1,0) state in Fig.~\ref{starkcurves}, using a traditional Stark decelerator. As only a limited electric field can be used, a long decelerator would be required. Unfortunately the coupling of longitudinal and transverse motion in a traditional Stark decelerator reduces the phase-space acceptance considerably, which is problematic for long decelerators. A possible solution, especially if one does not have to reach standstill, is to use a different switching mode in the decelerator~\cite{2005PhRvA..71e3409V,Meerakker:2006tx,Scharfenberg:2009hta}. For heavy molecules, however, this is not a solution, as the increased transverse stability is obtained by using the high electric fields in between the electrodes. Such high electric fields would drive the heavier molecules into the high-field seeking regime, which would lead to additional losses.

The purpose of the study presented here is to evaluate the performance of a traveling-wave Stark decelerator using ring-shaped electrodes (called ring-decelerator) for the deceleration and trapping of heavy diatomic molecules in weak-field seeking states. In a recent experiment with this type of decelerator the deceleration of CO has been demonstrated~\cite{2010PhRvA..81e1401O}. As was mentioned in reference~\cite{2010PhRvA..81e1401O}, a ring-decelerator is promising for heavy diatomic molecules because the molecules remain confined during the deceleration process in a trap with limited electric field strength. A second advantage is the stability of the deceleration process, which allows for the construction of efficient long decelerators. Finally, losses in the transfer from the decelerator to a separate trap~\cite{2010EPJD...57...33G} are avoided, because molecules can be stopped and trapped by the same electrodes that form the decelerator. We have picked the SrF molecule as the prototype molecule for our simulations, because is predicted to be a sensitive probe for molecular parity violation~\cite{Demille:2008he}, because it has a reasonable dipole moment (3.49 Debye), rotational moment and corresponding Stark shift and because is not too heavy so that it can still be decelerated in a reasonable size decelerator.

\section{Ring-deceleration of SrF molecules}
\subsection{Principle of operation}
\begin{figure}
\resizebox{\linewidth}{!}{\includegraphics{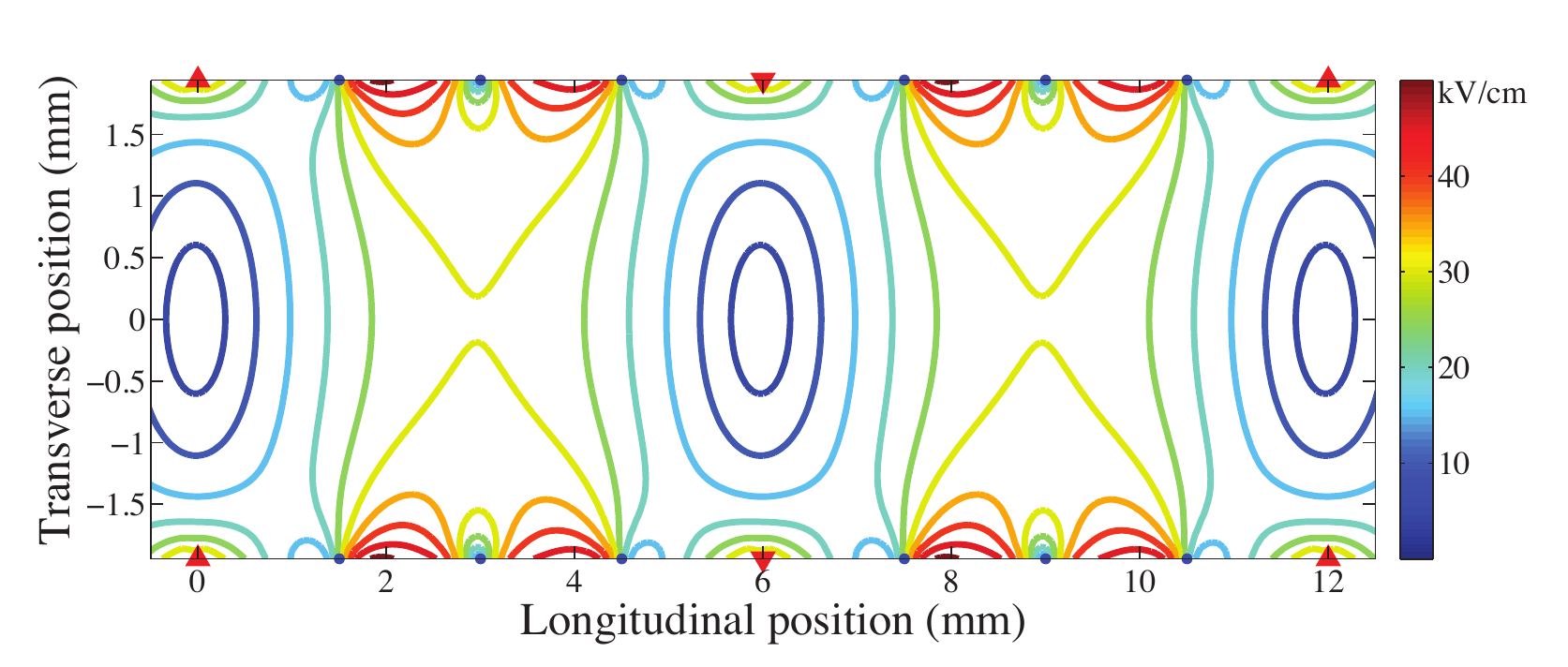}}
 \caption{\label{electricfield}Lines of equal electric field strength show the electric field minima in this longitudinal cut through a section of the decelerator. The blue dots at the top and the bottom mark the position of a ring electrode. To the electrodes marked with a red arrow pointing up and down the maximum and minimum potential is applied, respectively. The minima move through the decelerator with a velocity that is determined by the oscillation frequency of the potential on the electrodes.}
\end{figure}

Here we briefly summarize the operational principle of a ring-decelerator. For more details we refer to the existing literature on this type of decelerator~\cite{2010PhRvA..81e1401O, Meek:2009dg}. A large number of ring-shaped electrodes (depending on the length of the decelerator a few hundred to a few thousand) are connected in 8 sets to 8 high-voltage supplies. Oscillating voltages are applied to these sets of ring electrodes with a phase-difference of $2\pi / 8$, thereby creating a series of moving electric field minima in the decelerator. These minima are traps for molecules in a weak-field seeking state. In Fig.~\ref{electricfield} lines of equal electric field strength are drawn on a longitudinal cut through the decelerator. The electric field dependence of the Stark shift determines the shape of the trapping potential created by this electric field. For the typical heavy diatomic molecule SrF in the X$^{2}\Sigma^{+}(v=0)$ ground state the Stark shift is plotted in Fig.~\ref{starkcurves}. For comparison, also the Stark shift of CO in the a$^{3}\Pi_{1}$ state is given. The frequency of the applied voltage determines the velocity of the moving traps; the amplitude determines the depth of the traps. Initially, as the molecules enter the decelerator, the traps are set to move at the same speed as the molecules. Then, gradually, the oscillation frequency of the voltages is swept down resulting in the deceleration and ultimately stopping of the traps, with (a fraction of) the molecules remaining in the traps.

The deceleration of the traps can be added as a pseudo force to the force that the molecules experience in the frame of the moving traps. With increasing deceleration speed the addition of the deceleration force reduces the trap volume, until a maximum deceleration force is reached that is equal in strength to the trapping force. At this point the acceptance of the decelerator is reduced to zero. Therefore, to stop a molecule in a specific state with a given initial velocity, a minimum length of the decelerator is required. Increasing the length of the decelerator beyond this minimum will increase the fraction of decelerated molecules. An important difference between a ring-decelerator and a traditional Stark decelerator is the stability in long decelerators: we discuss this in more detail further on in the paper.

\subsection{Parameters and input for the simulations}
Using finite-element methods we have calculated the time- and position dependent electric field in the decelerator. Cylindrical symmetry is used to speed up the calculations. The electric fields are assumed to be periodic in the decelerator structure. We sample the movement of the traps due to an entire sinusoidal period in 80 frames. Edge-effects such as distortions of the electric field lines at the entrance of the decelerator are not present in the simulations but are expected to play a role in the experiment. We have not taken into account inter-molecular collisions and collisions with background gas, effects that we expect to play a minor role in the experiment. Using the Stark shift of the molecular state the electric field map is transformed to a potential, from which the force is derived. The trajectory of the molecules through the decelerator is then calculated numerically. By comparing to the experimental results~\cite{2010PhRvA..81e1401O} on the ring-deceleration of CO we could check the validity of our simulation code.

The inner diameter of the ring electrodes is 4 mm, the electrodes have a thickness of 0.6 mm, and the center-to-center spacing of the electrodes is 1.5 mm. These dimensions correspond to the ring-decelerator that has been used to decelerate CO~\cite{2010PhRvA..81e1401O}. The applied voltage to a set of electrodes is a sine wave with an amplitude that we have varied in the simulation over a range from 5 kV to 8 kV (peak-to-peak voltage of 10 to 16 kV). The availability of suitable high voltage supplies to operate the decelerator will determine the voltage that can be applied in actual experiments. Between each of the 8 sets of electrodes a phase-shift of 2$\pi$/8 is applied. We have calculated the deceleration process for various $(\mathrm{N,M}=0)$ states of the SrF molecule, denoted in the following text as SrF(N,0). The Stark shift of SrF is calculated with the aid of \textsc{pgopher}~\cite{Western:wl} and molecular parameters available from literature~\cite{Ernst:1985dp,Childs:1981ka,Le:2009kh,STEIMLE:1977td}. 

\subsection{Total 1D Phase-space acceptance}\label{subsec:totalacceptance}
As a first assessment of the performance of the decelerator for SrF we have calculated the phase-space acceptance of the decelerator, as a function of the applied voltage and deceleration strength. In this first step we have neglected 3D effects, further on in the paper we will discuss such effects of coupling between the longitudinal and transverse motion. The longitudinal 1D acceptance is calculated as the area within the phase-stable area, called the separatrix. Molecules within this area will oscillate longitudinally around the trap center and are kept together during the deceleration process. Molecules outside this area can not be decelerated in a phase-stable manner. A similar procedure is used to obtain the transverse 1D phase-space acceptance. The resolution for the 1D-separatrix is 0.1 m/s per axis and is limited by the discrete set of starting velocities used. The product of the areas in phase-space that lie within the region bound by the three 1D separatrices we call the total 1D phase-space acceptance.

\subsubsection{Guiding}
\begin{figure}[h]
\centering
  \includegraphics[width=1\columnwidth]{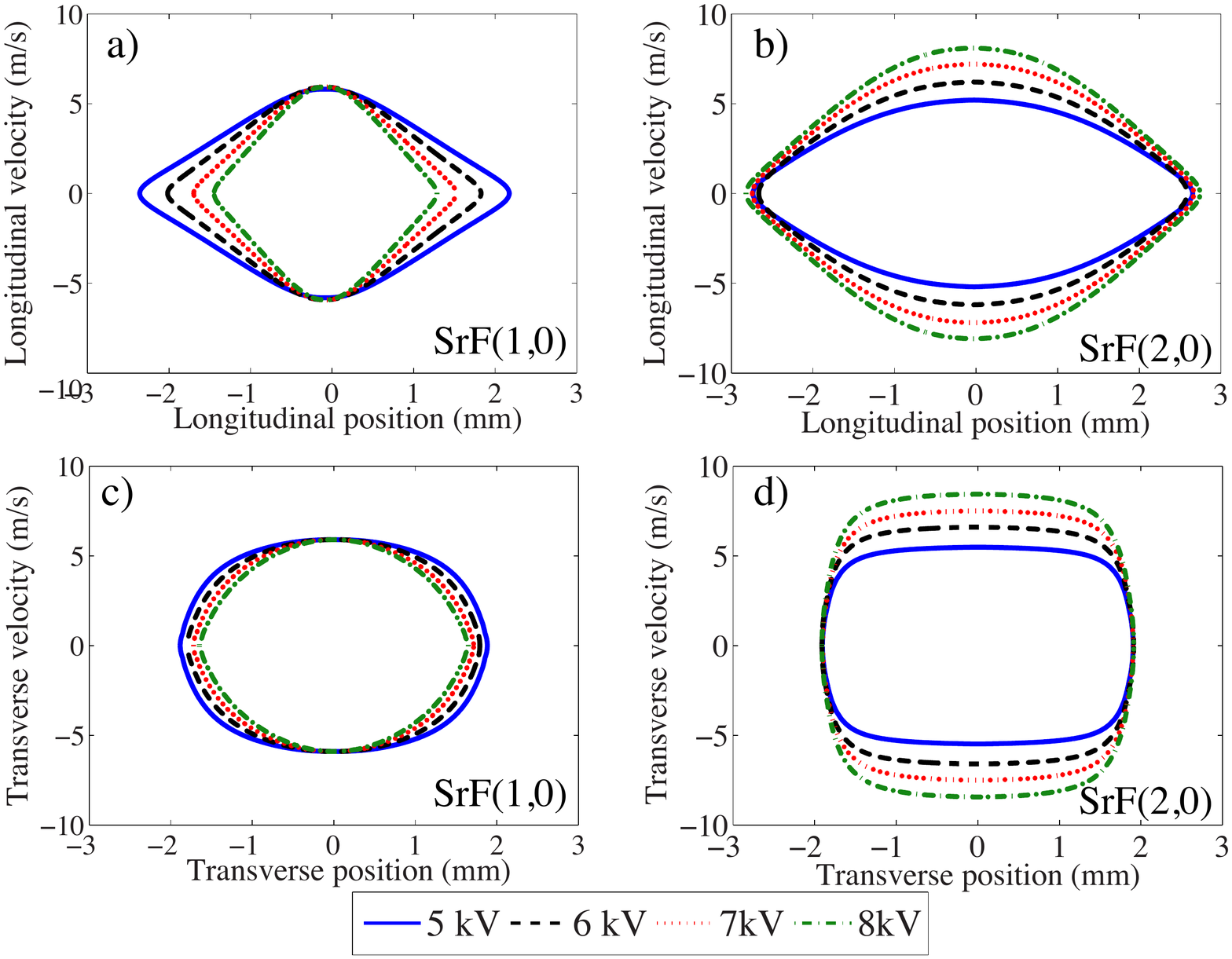}
  \caption{Longitudinal and transverse 1D separatrices for SrF(1,0) and SrF(2,0) for guiding, for four different applied voltages.}
  \label{guiding4box}
\end{figure}

The guiding of molecules at constant velocity is the mode of operation of the decelerator with the highest phase-space acceptance. The longitudinal and transverse separatrices for the guiding of SrF(1,0) and SrF(2,0) are plotted for different applied voltages in Fig.~\ref{guiding4box}. The traps, with a size of about 4x4 mm, can typically confine molecules with a velocity of between 5 and 10 m/s relative to the guiding traps. For SrF(1,0) the peculiar situation arises that for \textit{increasing} voltage (in the range of 5 to 8 kV) the acceptance \textit{decreases}, which is a consequence of the particular shape of the Stark curve. The result is a spatial compression of the trap for higher voltages. For guiding with the chosen geometry 5 kV has the highest acceptance, as for 4 kV (not shown) the maximum of the Stark curve is not yet reached. For SrF(2,0) the situation is reversed, as can be seen in Fig.~\ref{guiding4box}(b) and \ref{guiding4box}(d). For this state the maximum of the Stark shift has not yet been reached at the applied voltages; therefore the acceptance volume keeps growing with increasing voltage. There is a small dependence of the 1D acceptance on the position relative to the electrodes in the decelerator. This can lead during deceleration to a slightly lower acceptance than estimated here.

\subsubsection{Decelerating}
\begin{figure}[h]
\centering
  \includegraphics[width=1\columnwidth]{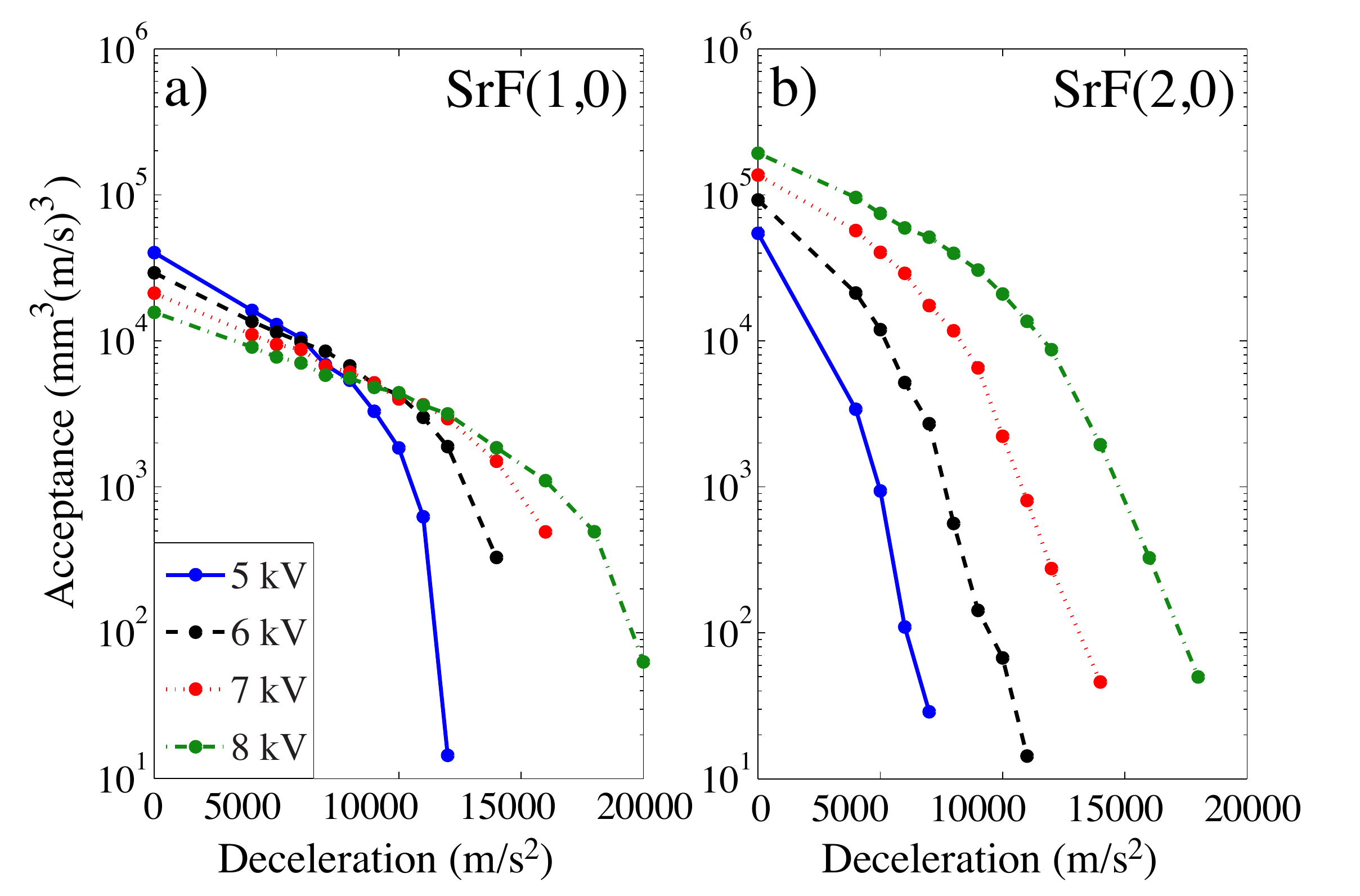}
  \caption{The total 1D acceptance of the ring-decelerator for (a) SrF(1,0) and (b) SrF(2,0), as a function of deceleration strength, for different applied voltages. In this calculation there was no coupling between the longitudinal and transverse motion.}
  \label{totalacceptance}
\end{figure}

When decelerating the acceptance is decreased compared to guiding. The amount depends on the deceleration strength, the state of the molecule, and the applied voltage. For a range of deceleration strengths we have calculated the total 1D phase-space acceptance, for both the SrF(1,0) and SrF(2,0) states. The results are summarized in Fig.~\ref{totalacceptance}. For SrF(1,0) it can be seen that with increasing deceleration strength the initially relatively high acceptance for low voltages is reduced, until around a deceleration of 8000 m/s$^{2}$ the total acceptance for deceleration of SrF(1,0) is practically independent of the applied voltage. This can be understood from the steeper slope of the trapping potential (at the cost of a smaller trap volume) for the higher voltages. For SrF(2,0) the situation is simpler. At low deceleration strengths and high voltages the acceptance for SrF(2,0) is about 10 times larger than for SrF(1,0), whereas for the very high deceleration strengths SrF(1,0) has a slightly higher acceptance. This difference in the acceptance at very high deceleration strengths between the SrF(1,0) and SrF(2,0) level reflects the difference in the slope of the Stark curves.

With a deceleration strength of 9000 m/s$^{2}$ a beam of SrF molecules with an initial velocity of 300 m/s (which can be reached in a cooled supersonic expansion using xenon as a carrier gas) can be stopped using a 5 meter long decelerator. This length is similar to that needed for alternative deceleration approaches. The typical length of traditional Stark decelerators for light molecules such as OH, NH$_{3}$ and CO is around 1 meter.

\subsection{Total 3D phase-space acceptance}\label{coupling}
\begin{figure}[h]
\centering
  \includegraphics[width=1\columnwidth]{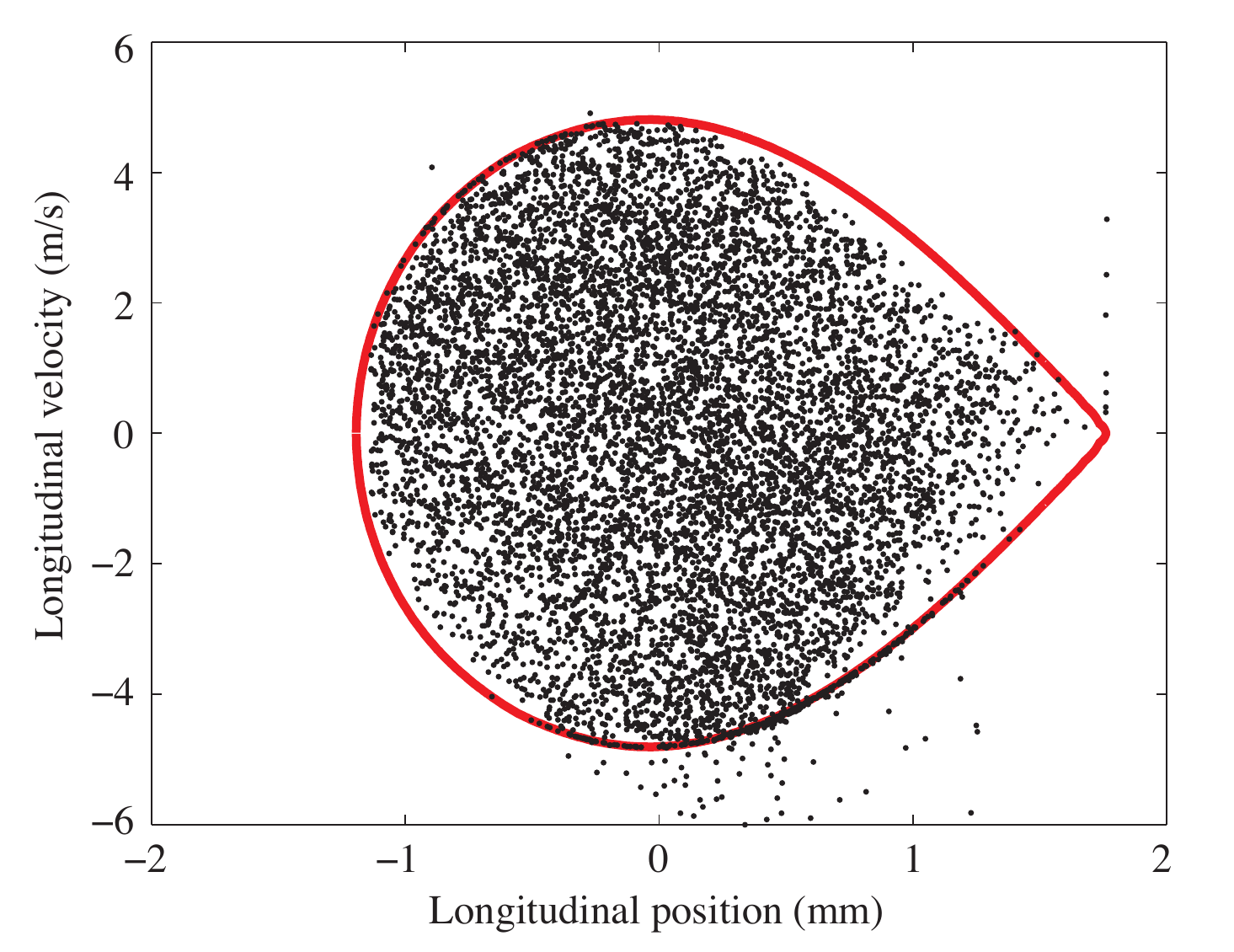}
  \caption{The 1D longitudinal separatrix for the deceleration with 9000 m/s$^{2}$ of SrF(2,0) at 8 kV (red solid line), with the result of a full 3D simulation of 50000 molecules (black dots). The fact that almost the complete 1D separatrix is uniformly filled by the particles illustrates the stability of the deceleration process in the ring-decelerator.}
  \label{molinsep}
\end{figure}

In this section we present the full 3D calculation of the deceleration of SrF molecules using a ring-decelerator. It has been shown for traditional Stark decelerators~\cite{Meerakker:2006tx,Scharfenberg:2009hta} that molecules are lost from within the 1D separatrices during the deceleration process due to coupling of the longitudinal and transverse motion. This effect leads to larger losses for longer decelerators. There is also a minimum length required to stop a reasonable fraction of the molecules. For traditional decelerators there is therefore an optimum length to reach a desired final velocity. A furthermore especially critical part is around low velocities where the molecules are transversely over-focused and hit the electrodes. The situation for a ring-decelerator is quite different, as is illustrated by Fig.~\ref{molinsep}. Here the result of a full 3D calculation is compared to the calculated 1D longitudinal acceptance, for the SrF(2,0) state with a deceleration strength of 9000 m/s$^{2}$ and 8 kV applied voltage. For every datapoint in the 3D simulations we used 5000 molecules distributed over the 1D separatrix acceptance. The results of the 3D acceptance calculations have a statistical scatter of $5~\textrm{-} ~10\%$.

Over 75\% of the molecules that started within the total 1D acceptance make it to the end in this 5 meter long decelerator. Some molecules are lost through collisions with the ring-shaped electrodes, or escape longitudinally over the potential barrier. The small phase-space reduction caused by the movement of the trap as it moves along the electrodes clearly does not lead to large losses. This high stability is a key issue for the good performance of long ring-decelerators for heavy diatomic molecules.

\begin{figure}[h]
\centering
  \includegraphics[width=1\columnwidth]{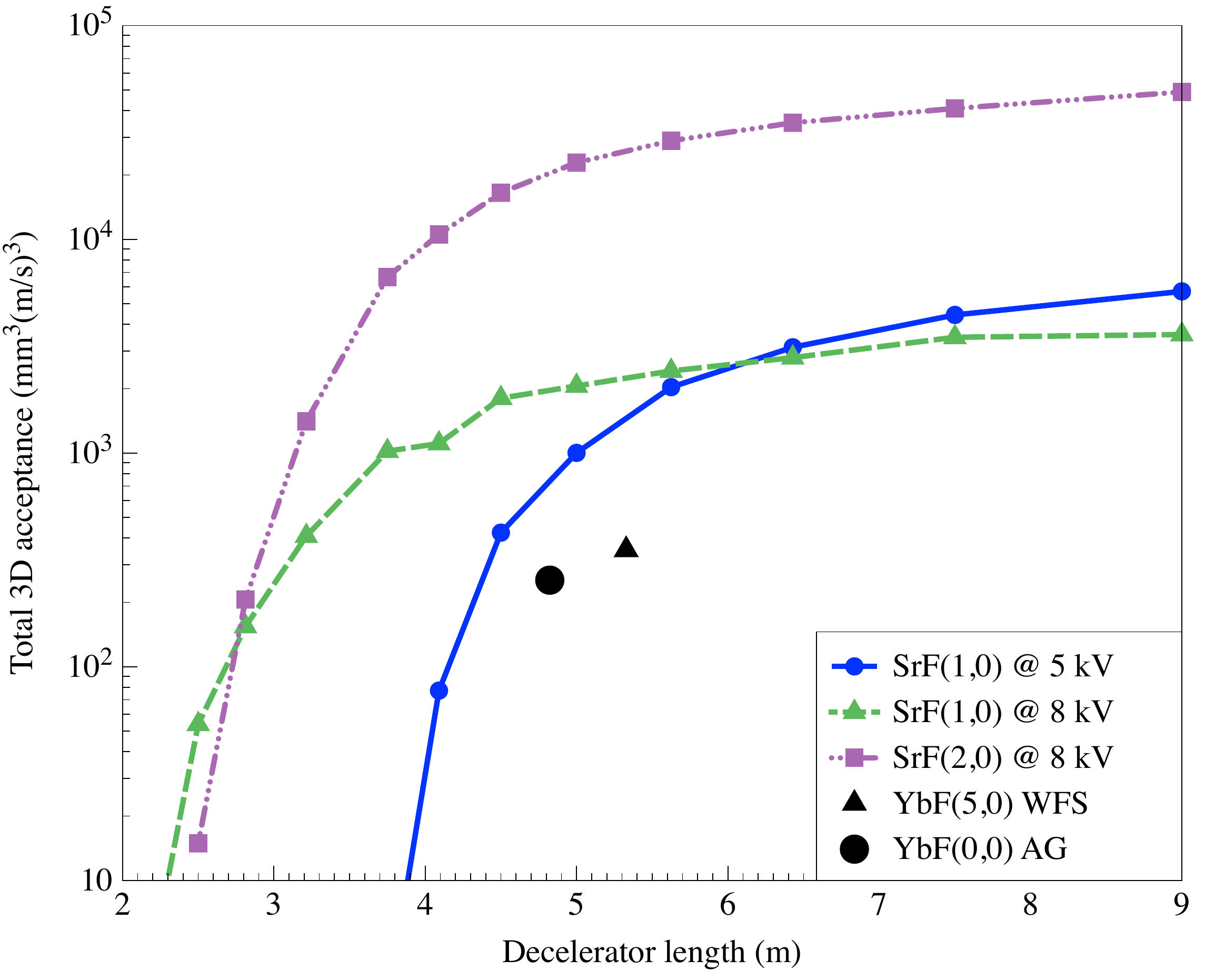}
  \caption{The total 3D acceptance of the ring-decelerator for SrF(1,0) and SrF(2,0) as a function of the total length of the decelerator, for two different applied voltages. The deceleration strength is adjusted to stop and trap the molecules from an initial 300 m/s in the given length. The two points denote calculation results~\cite{Tarbutt:2009cf} for other deceleration methods, which are discussed in the text.}
  \label{nmol-vs-decel}
\end{figure}

In Fig.~\ref{nmol-vs-decel} the total 3D acceptance of the ring-decelerator for SrF(1,0) and (2,0) is shown, as a function of decelerator length, for 5 kV and 8 kV. For a fixed initial velocity of 300 m/s the deceleration strength is varied, leading to a different decelerator length in order to reach a standstill. This is the result of a full 3D simulation, allowing for coupling between the longitudinal and transverse motion. We uniformly and randomly fill the total 1D acceptance volume at the beginning of the simulation, and then track the fraction of molecules that remain within this volume. The total acceptance that is shown on the vertical axis is the product of the total 1D acceptance as it is plotted in Fig.~\ref{totalacceptance} and the fraction of molecules that remain within the 1D separatrices throughout the full 3D calculation. The performance for an applied voltage of 8 kV is clearly superior to 5 kV, especially when using SrF(2,0). Comparing SrF(2,0) at 8kV to SrF(1,0) at 5 kV, the same acceptance can be reached for a 3 meter long decelerator instead of a 5 meter long decelerator. If a 5 meter long decelerator would be used at 8 kV for SrF(2,0), the total acceptance is 20 times higher than for 5 kV SrF(1,0).

\section{Comparison to other deceleration approaches} \label{approaches}
\begin{figure}[h]
\centering
  \includegraphics[width=1\columnwidth]{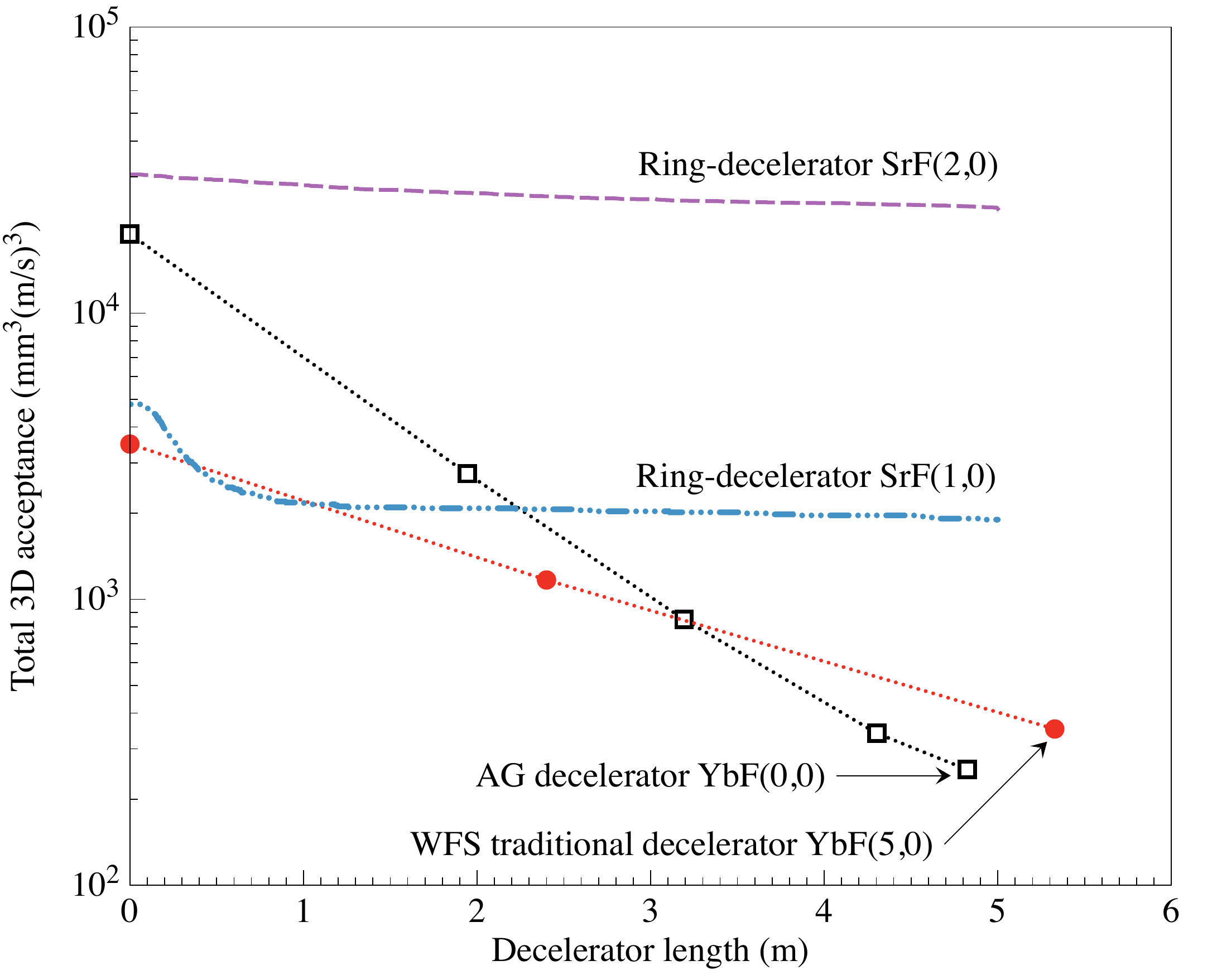}
  \caption{A comparison of the decrease of the calculated total phase-space acceptance during the deceleration process for different types of decelerators for heavy diatomic molecules. The alternate gradient (AG) deceleration data is for YbF in the (0,0) state, the weak-field seeking (WFS) traditional deceleration is for YbF in the (5,0) state. For both decelerators the total length required to decelerate the molecules to a velocity that can be trapped is about 5 m, the initial velocity is 340 m/s, and the applied voltages are $\pm$~40 kV. The data for these simulations is taken from Tarbutt \textit{et al}~\cite{Tarbutt:2009cf}. For the ring-decelerator the acceptance for the SrF(1,0) and SrF(2,0) states at $\pm$8~kV is shown.}
  \label{comparingdecel}
\end{figure}

In this section we compare the results on ring-deceleration that we presented in the previous paragraphs to the two different approaches that can be taken to decelerate heavy diatomic molecules: alternate gradient (AG) deceleration and traditional Stark deceleration in a weak-field seeking (WFS) excited rotational state. Since no heavy molecules have been trapped yet, we rely on the results of simulations. The two alternative methods have been compared for YbF molecules by Tarbutt \textit{et al}~\cite{Tarbutt:2009cf}. The data from their simulations is plotted in Fig.~\ref{comparingdecel}. It shows the decrease in the number of molecules for a given deceleration strength as they travel through the decelerator. The deceleration strength is sufficient to reach a low enough velocity that the molecules could be loaded into a trap. The final two datapoints are also included in Fig.~\ref{nmol-vs-decel}. For the WFS deceleration a traditional decelerator with a gap of 4 mm square and an applied voltage of $\pm$ 40 kV was assumed, for the AG deceleration the same voltage difference was used. For the WFS simulation the YbF(5,0) molecules were taken; since normally in the supersonic expansion this state is barely populated special microwave pumping schemes would have to be used before entering the decelerator. For both deceleration techniques about 5 meters of decelerator is required to bring the molecules to a sufficiently low velocity such that they can be trapped.

The data from the simulation of the ring-deceleration of SrF is also plotted in Fig.~\ref{comparingdecel}. It is hard to make a direct quantitative comparison between the different deceleration methods, as the performance of the decelerators depends very strongly on the electrode geometry and decelerator length, the applied voltages, the molecule and which state is used. This can be seen in Fig.~\ref{nmol-vs-decel}. Nevertheless, it is clear from these figures that the ring-decelerator can outperform these alternative methods by at least an order of magnitude while using much lower electric fields. The main reason for this is that during the deceleration process a much larger fraction of the molecules remains trapped, due to the high phase-space stability in the ring-decelerator.

\section{\label{trapping}Trapping}
At the end of the deceleration process the moving trap, including the molecules, is brought to a standstill. The molecules remain trapped in the last section of the decelerator, where further experiments can be performed. As a result there are no additional losses of molecules in the transfer from the decelerator to the trap. For traditional decelerators this step can lead to substantial (factor 4-6) losses~\cite{2010EPJD...57...33G}.
For heavy diatomics the depth of the trap is limited by the maximum energy at the turning point of the Stark curve: for SrF(1,0) this is 0.15 cm$^{-1}$. A SrF molecule with this maximum kinetic energy moves at 5.8 m/s, which is the most probable velocity for a gas at a temperature of 215 mK. Operating the trap at higher voltages will, due to the special shape of the Stark curve for SrF in the (1,0) state, not lead to a deeper trap, but to a trap with the same depth and a reduced volume.
To further reduce the temperature either sympathetic cooling (which has so far only been analyzed for lighter systems~\cite{Zuchowski:2008eq,Soldan:2009ig,Tokunaga:2011dla}) or direct laser cooling techniques could be explored. Recently laser cooling of SrF has been shown~\cite{Shuman:2009ja,Shuman:2010gpa}, in the first demonstration of the laser cooling of a molecule. We would further like to remark that the effect of blackbody radiation on the trapped molecules~\cite{Hoekstra:2007dk,Vanhaecke:2008jh} is only significant on timescales larger than one second.

\section{Measurement of parity violation using decelerated molecules}
A parity violation measurement in molecules using a Stark-interference technique was proposed by DeMille~\cite{Demille:2008he}. The statistical uncertainty of such an experiment scales as $\sigma \sim 1/\sqrt{N}\tau$, where $N$ denotes the number of molecules and $\tau$ the coherent measurement time. 

To compare the flux of molecules with a molecular beam experiment we first estimate the number of trapped molecules. As the SrF source is very similar to the YbF source reported on in reference \cite{Tarbutt:2002wv}, we assume that the same ground state beam flux of $1.4 \times 10^9$ molecules/sr can be achieved. Convoluting this with the decelerator acceptance we expect to start the deceleration process with $\sim 10^5$ SrF(1,0) molecules per shot.  As can be seen from Fig.~\ref{totalacceptance}, when decelerating at $9000  \textrm{ m/s}^2$ the acceptance is reduced compared to guiding. Also during the deceleration some molecules are lost. In total this leads to a deceleration efficiency of $\sim 10 \%$. Since no more molecules are lost upon trapping, we expect to achieve $\sim 10^4$ trapped SrF(1,0) per shot. While operating in a 10 Hz pulsed mode this corresponds to a flux $N = 10^5/\textrm{s}$. This number is therefore comparable to the proposed cryogenic molecular beam experiment~\cite{Demille:2008he}. 

We now turn to the potential gain in interaction time. Since the environment of an electrostatic trap introduces a number of challenges for precision experiments, the use of slow and/or laser-cooled (but not trapped) molecules is currently the most promising approach. In a molecular beam experiment the interaction time $\tau$ is limited by the longitudinal velocity of the beam and the length of the interaction region that it traverses. For a typical cryogenic beam velocity of 150 m/s and a measurement region of $50 \times 5 \times 5$ mm~\cite{Demille:2008he}, this results in $\tau = 0.3 \textrm{ ms}$. The transverse velocity determines the size of the molecular beam at the exit of the measurement region. For a measurement region length of 50 mm and a beam with $v_{long} \ge 10 \times v_{trans}$, the transverse size increase of the beam is limited to $\le 5$ mm. As pointed out in the previous paragraph, the decelerated packets of molecules have a speed of $\le 6$ m/s. To make use of these molecules while keeping them in the $50 \times 5 \times 5$ mm interaction region we would have to give the packet of molecules a longitudinal velocity of $\sim 60$ m/s. This would result in a modest gain in the interaction time of a factor of three. Clearly, one can only make significant gains in the interaction time using molecular beams if the transverse velocity is also significantly reduced. To achieve this one can use laser cooling on the decelerated molecules, by which the temperature can be reduced to $\sim 150 ~\mu$K. At this temperature the most probably velocity drops to 0.17 m/s, indicating that a longitudinal velocity of 1.7 m/s would be sufficient to keep the beam within the above mentioned size criteria, resulting in an interaction time increase by a factor $\sim 90$. 

\section{Deceleration of other molecules}
Following this analysis of the deceleration of SrF molecules in a 3 to 5 meter long ring-decelerator, we turn to an assessment of the deceleration of other relevant molecules. The alkaline-earth monohalides CaF, BaF and also YbF all have an X$^{2}\Sigma^{+}$ groundstate with a similar Stark shift as SrF. For the (1,0) states we assume that a sufficiently high voltage can be applied such that the maximum of the Stark curve is limiting the acceptance. In that specific case the decelerator length is independent of the electric dipole moment of the molecule. Therefore, the required length of the decelerator to reach a comparable deceleration efficiency as for SrF(1,0) is proportional to the ratio of the mass and the rotational constant. For RaF we do not have the complete information to make this estimation~\cite{2010PhRvA..82e2521I}. The results are given in Table~\ref{othermolecules}.

\begin{table}[h]
\small
  \caption{\ Ring-deceleration of other alkaline-earth monohalides}
  \label{othermolecules}
  \begin{tabular*}{0.5\textwidth}{@{\extracolsep{\fill}}lllll}
    \hline
    Molecule & Mass & Dipole moment & Rotational & Relative\\
     & (amu) & (Debye) & constant (cm$^{-1}$) & length \\
    \hline
    CaF(1,0) & 59 &3.07 & 0.37 & 0.38\\
    SrF(1,0) & 106 &3.49 & 0.25 & 1\\
    BaF(1,0) & 156 &3.17 & 0.21 &  1.75\\
    YbF(1,0) & 192 &3.91 & 0.24 &  1.89\\
    \hline
  \end{tabular*}
\end{table}

The deceleration is not restricted to molecules in X$^{2}\Sigma^{+}$ states. For example, also the deceleration of PbF in the X$^{2}\Pi$ state looks promising, as a Stark shift of $\sim$ 0.3 cm$^{-1}$ is reached at moderate electric fields of 15 kV/cm. And finally we point out that also for the lighter molecules a long ring-decelerator would compare favorably to existing Stark decelerators.
 
\section{Conclusions and outlook}
In this paper we have presented a quantitative study of the deceleration and trapping of heavy diatomic molecules using a ring-decelerator. Trapped samples of such molecules are very promising for the study of the violation of fundamental discrete symmetries. For a prototypical selected molecule, SrF, it is shown that a 5 meter long decelerator with modest applied voltages (amplitude of 8 kV) has a total phase-space acceptance for deceleration in the second excited rotational state of about $2 \times 10^4$ mm$^{3}$(m/s)$^{3}$. This is at least an order of magnitude larger compared to alternative deceleration methods using a similar decelerator length and deceleration strength, but operating at much higher voltages ($\pm$40 kV). The construction of the analyzed ring-decelerator is currently underway at our institute, in collaboration with Rick Bethlem (VU University Amsterdam).

We have estimated the potential gain in sensitivity of using the decelerated molecules for a parity violation measurement. Compared to a cryogenic molecular beam experiment, the decelerated molecules provide a gain in sensitivity of up to two orders of magnitude due to the increased interaction time, while maintaining a comparable molecular flux. By combining this efficient form of Stark deceleration with a final stage of laser cooling a robust method to create ultracold trapped samples of heavy diatomic molecules is within reach.

\begin{acknowledgements}
We acknowledge fruitful discussions with Klaus Jungmann, Ronnie Hoekstra, Timur Isaev, Robert Berger and Rick Bethlem, and helpful suggestions by the referees. These experiments are embedded within the Fundamental Interactions and Symmetries research line at the KVI. This work is part of the research programme of the Foundation for Fundamental Research on Matter (FOM), which is part of the Netherlands Organisation for Scientific Research (NWO) (FOM Program nr. 125). 
\end{acknowledgements}
\footnotesize{
\bibliography{$HOME/documents/bib/bib} 
}

\end{document}